\documentclass[final,5p,times,twocolumn]{elsarticle}

\usepackage{amssymb}
\usepackage{lipsum}

\usepackage{cleveref}

\usepackage{lineno}
\setcitestyle{square}

\journal{Nuclear Instruments and Methods in Physics Research A}

\begin{document}

\begin{frontmatter}



\title{Refractive index in the JUNO liquid scintillator}

\author[ihep,ucas]{H.S.~Zhang}
\author[b]{M.\,Beretta\corref{cor1}}\ead{marco.beretta@mi.infn.it}
\author[b]{S.\,Cialdi}
\author[ihep,ustc]{C.X.~Yang}
\author[ihep,ucas]{J.H.~Huang}
\author[b]{F.\,Ferraro}
\author[ihep,ucas]{G.F.~Cao\corref{cor1}}\ead{caogf@ihep.ac.cn}
\author[b]{G. Reina}
\author[ihep]{Z.Y.~Deng}
\author[b]{E. Suerra}
\author[b]{S. Altilia}
\author[b]{V. Antonelli}
\author[b]{D. Basilico}
\author[b]{A. Brigatti}
\author[b]{B. Caccianiga}
\author[b]{M.G. Giammarchi}
\author[b]{C. Landini}
\author[b]{P. Lombardi}
\author[b]{L. Miramonti}
\author[b]{E. Percalli}
\author[b]{G. Ranucci}
\author[b]{A.C. Re}
\author[b]{P. Saggese}
\author[b]{M.D.C. Torri}
\author[a]{S. Aiello}
\author[a]{G. Andronico}
\author[e]{A. Barresi}
\author[c]{A. Bergnoli}
\author[e]{M. Borghesi}
\author[c]{R. Brugnera}
\author[a]{R. Bruno}
\author[d]{A. Budano}
\author[j]{A. Cammi}
\author[c]{V. Cerrone}
\author[a]{R. Caruso}
\author[e]{D. Chiesa}
\author[f]{C. Clementi}
\author[c]{S. Dusini}
\author[d]{A. Fabbri}
\author[g]{G. Felici}
\author[c]{A. Garfagnini}
\author[a]{N. Giudice}
\author[c]{A. Gavrikov}
\author[c]{M. Grassi}
\author[c]{R.M. Guizzetti}
\author[a]{N. Guardone}
\author[c]{B. Jelmini}
\author[c]{L. Lastrucci}
\author[c]{I. Lippi}
\author[j]{L. Loi}
\author[a]{C. Lombardo}
\author[h,i]{F. Mantovani}
\author[d]{S.M. Mari}
\author[g]{A. Martini}
\author[h,i]{M. Montuschi}
\author[e]{M. Nastasi}
\author[d]{D. Orestano}
\author[f]{F. Ortica}
\author[g]{A. Paoloni}
\author[d]{F. Petrucci}
\author[e]{E. Previtali}
\author[c]{M. Redchuck}
\author[h,i]{B. Ricci}
\author[f]{A. Romani}
\author[a]{G. Sava}
\author[c]{A. Serafini}
\author[c]{C. Sirignano}
\author[e]{M. Sisti}
\author[c]{L. Stanco}
\author[d]{E. Stanescu Farilla}
\author[h,i]{V. Strati}
\author[c]{A. Triossi}
\author[a]{C. Tuve'}
\author[d]{C. Venettacci}
\author[a]{G. Verde}
\author[g]{L. Votano}

\cortext[cor1]{Corresponding authors}
\address[ihep]{Institute of High Energy Physics, Chinese Academy of Sciences, Beijing 100049, China}
\address[ucas]{University of Chinese Academy of Sciences, Beijing 100049, China}
\address[ustc]{University of Science and Technology of China, Hefei 230026, China}
\address[b]{INFN, Sezione di Milano e Università degli Studi di Milano, Dipartimento di Fisica, Italy}
\address[a]{INFN, Sezione di Catania e Università di Catania, Dipartimento di Fisica e Astronomia, Italy}
\address[c]{INFN, sezione di Padova e Università di Padova, Dipartimento di Fisica e Astronomia, Italy}
\address[d]{INFN, sezione di Roma Tre e Università degli Studi Roma Tre, Dipartimento di Fisica e Matematica, Italy}
\address[e]{INFN, Sezione di Milano Bicocca e Dipartimento di Fisica Università di Milano Bicocca, Italy}
\address[f]{INFN, Sezione di Perugia e Università degli Studi di Perugia, Dipartimento di Chimica, Biologia e Biotecnologie, Italy}
\address[g]{Laboratori Nazionali dell’INFN di Frascati, Italy}
\address[h]{INFN, Sezione di Ferrara, Italy}
\address[i]{Università degli Studi di Ferrara, Dipartimento di Fisica e Scienze della Terra, Italy}
\address[j]{INFN, Sezione di Milano Bicocca e Dipartimento di Energetica, Politecnico di Milano, Italy}

\begin{abstract}
In the field of rare event physics, it is common to have huge masses of organic liquid scintillator as detection medium. In particular, they are widely used to study neutrino properties or astrophysical neutrinos. Thanks to its safety properties (such as low toxicity and high flash point) and easy scalability, linear alkyl benzene is the most common solvent used to produce liquid scintillators for large mass experiments. The knowledge of the refractive index is a pivotal point to understand the detector response, as this quantity (and its wavelength dependence) affects the Cherenkov radiation and photon propagation in the medium. In this paper, we report the measurement of the refractive index of the JUNO liquid scintillator between 260-1064 nm performed with two different methods (an ellipsometer and a refractometer), with a sub percent level precision. In addition, we used an interferometer to measure the group velocity in the JUNO liquid scintillator and verify the expected value derived from the refractive index measurements. 
\end{abstract}



\begin{keyword}
Refractive index \sep linear alkyl benzene \sep Cherenkov light \sep scintillator \sep neutrino physics \sep JUNO experiment



\end{keyword}

\end{frontmatter}




\section{Introduction}
\label{introduction}

The linear alkyl benzene, also known as LAB, is one of the most common solvents to produce modern organic liquid scintillators. For example, the Daya Bay experiment used 20x8 tons of LAB doped with Gadolinium to measure  the $\theta_{13}$ neutrino oscillation parameter with high accuracy \cite{DBLS}. Thanks to its favorable safety properties (low toxicity and high flash point), large scalability (a common solvent in the industry), and the possibility of being purified successfully, LAB will be used in several future neutrino experiments, like JUNO~\cite{JUNO} or SNO+~\cite{SNO} .
	
In particular, the JUNO experiment is a huge neutrino detector, under construction in China, which will use 20 ktons of liquid scintillator, based on LAB, contained in a 35 m diameter acrylic sphere. The primary scientific goal of JUNO is the determination of the neutrino mass ordering which, according to detailed sensitivity studies, is expected to require six years of data taking.
	
To reach this challenging task, JUNO will need an unprecedented energy resolution for a liquid scintillator based detector aiming to obtain $3\%$ at 1 MeV and a position resolution of $7\,\mathrm{cm}$ at 1~MeV~\cite{JUNO}. 
In order to reach the required performances on energy and position reconstruction, it is mandatory to know with high accuracy all the optical properties of the scintillator. In particular, this paper is focused on the refractive index of the JUNO scintillator.
In this paper, the results on the measurement of the refractive index, together with the two techniques that were used to obtain these results in different wavelength intervals, are presented.
	
	
The refractive index determines the velocity of photons in the medium. This is a crucial input for the reconstruction of the event interaction vertex, which is based on the arrival times of each photon produced in a given event to each phototube. The position reconstruction is fundamental in the JUNO analysis for several reasons: first of all, it allows the selection of an innermost region of the detector (Fiducial Volume) where the external background is negligible due to the self-shielding of the scintillator. Secondly, it is crucial to select a couple of events spatially correlated due to the interaction of the reactor anti-neutrinos in the detector\footnote{ Reactor anti-neutrinos mostly interact via the Inverse Beta Decay (IBD) reaction, in which a positron and a neutron are produced via the anti-neutrino capture by a proton. The two emitted particles are time and spatial correlated hence looking at this correlation it is possible to reduce the background contribution.}. In addition the refractive index is connected to other properties of the liquid scintillator like the Rayleigh scattering length \cite{R1}\cite{R2}, which can spoils the energy resolution if not well controlled.
For these reasons, knowing the refractive index over a large range of wavelengths and with high accuracy is very important. Recently, the RENO collaboration measured the refractive index of six wavelengths of LAB and PC ~\cite{Yeo_2010}. And H. Wan Chan Tseung et al. used the ellipsometer to measure the refractive index of LAB-OPP in the range of 210~nm-1000~nm with high accuracy ~\cite{Wan_2011}. In this study, the 260~nm-1064~nm refractive index of LAB is measured by ellipsometer and refractometer.
	
The refractive index also has an impact on the capability to reconstruct the energy of events in JUNO. The reconstruction of the energy of particles interacting in JUNO is possible thanks to the collection of photons generated in its active volume. Together with scintillation photons, which are the majority, also Cherenkov photons are produced following the well-known Frank-Tamm formula:
\begin{equation}
      \frac{d^2N}{dxd\lambda} = \frac{2\pi e^2}{\lambda^2} \left(1 - \frac{1}{\beta^2n^2(\lambda)}\right)
\end{equation}
hence, an accurate knowledge of the refractive index will help to constrain the number of Cherenkov photons and, therefore, to precisely determine the energy response of the detector. 
	
Furthermore, the accurate knowledge of the Cherenkov contribution to the total detected light is important when the Correlated and Integrated Directionality (CID) method is applied. This method was originally developed by the Borexino collaboration \cite{CID}, to increase the sensitivity to solar neutrino \cite{Solar}.

\section{Refractive index measurements}
To measure the refractive index we used two different techniques covering different parts of the spectrum. The region from 260~nm to 500~nm was covered using an ellipsometer, located at the Institute of High Energy Physics (IHEP) in Beijing, allowing to determine the refractive index in a spectral region in which the LAB absorbs the incident light but still some light is reflected. From 400~nm to 1064~nm, we used a refractometer, built in the Physics department of the University of Milan, which allows to cross-check the measurements in the region between 400 nm and 500 nm and extends the measurements up to 1064 nm. 

The JUNO liquid scintillator is composed of LAB + 2.5 g/L PPO + 3 mg/L bis-MSB. Since the concentration of the fluor and wavelength shifter are very low, we assume that the refractive index of the liquid scintillator is dominated by the LAB one and we performed our measurements on the pure LAB. We checked our assumption for $\lambda=405\,\mathrm{nm}$ and verified its validity.

\subsection{Ellipsometric measurements}

Spectroscopic ellipsometry utilizes the polarization of reflected light on a sample to probe the dielectric properties of the material under investigation. In ellipsometric measurements, the primary focus is on quantifying the polarization of reflected light, thus mitigating concerns regarding the absorption of the light by the medium. This facilitated precise measurement of the refractive index in the vacuum ultraviolet region, in contrast to refractometric measurements, which are constrained by the need for the incident beam to pass through the sample, resulting in limitations due to absorbance.
 
\begin{figure}[h]
    \centering
    \includegraphics[width=0.45\textwidth]{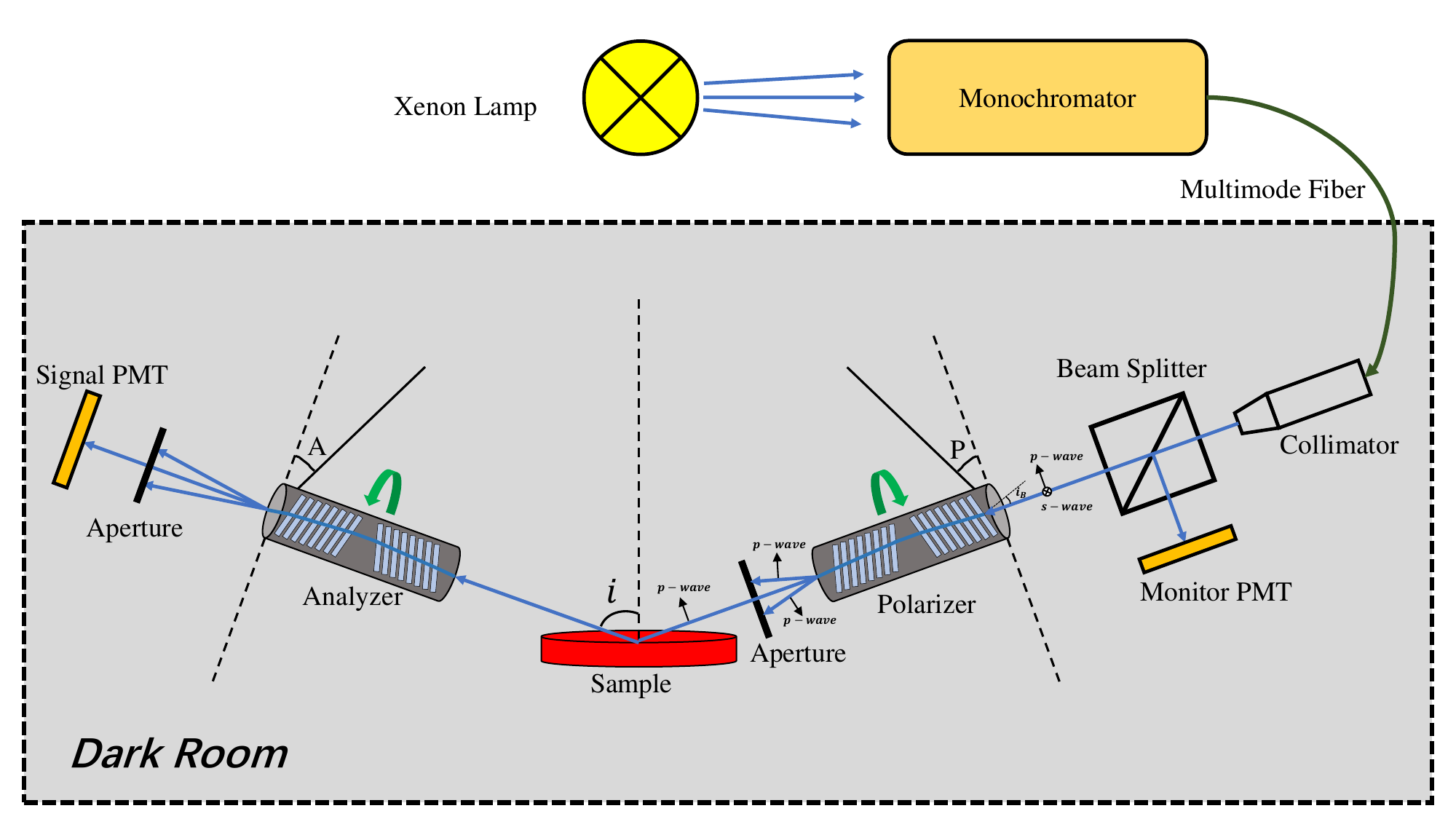}
    \includegraphics[width=0.45\textwidth]{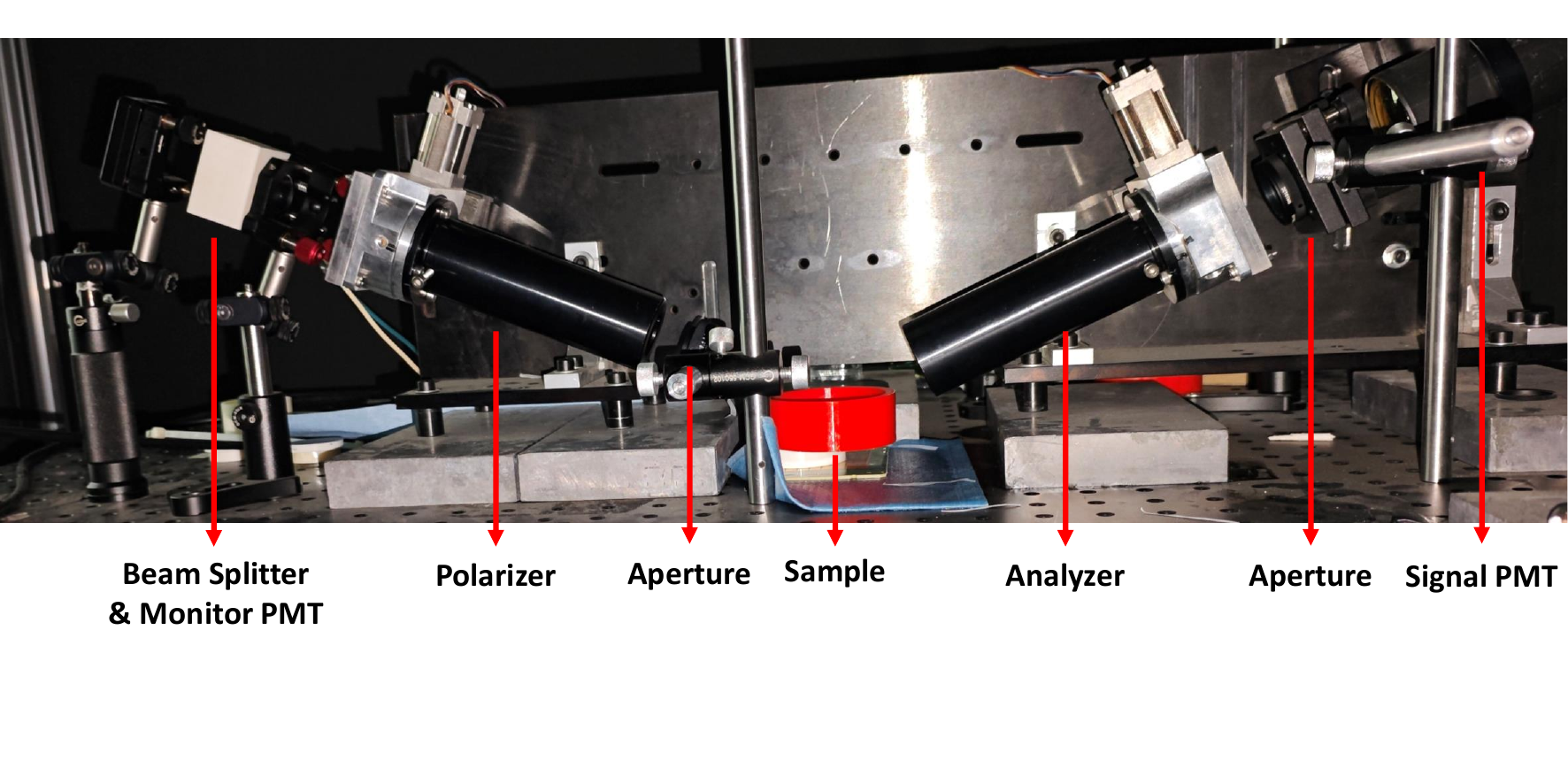}
    \caption{Top: A schematic diagram of the experimental setup for the ellipsometric measurements. Bottom: A photograph of the setup in the dark room.}
    \label{fig:ellipsometer}
\end{figure}

Figure~\ref{fig:ellipsometer} shows a schematic diagram and a photograph of the experimental setup for the polarization measurements. The light emitted from a xenon lamp is directed through a monochromator to select the wavelength of the incident radiation. Subsequently, the light is collimated and directed through a beam splitter, with a PMT (monitor PMT) positioned to correct the potential instabilities of the xenon lamp. The incident light is polarized by a polarizer and directed onto the sample through an aperture. The polarization of the reflected light is examined via an analyzer (a second polarizer) and a PMT (signal PMT). The polarizer and analyzer are both constructed with 16 calcium fluoride (CaF$_2$) windows, organized in parallel stacks of 8 pieces, and positioned symmetrically to prevent any displacement of the beam's center as the polarizer rotates. As the light passes through each CaF$_2$ window at the Brewster angle, the s-mode polarized light is diminished by reflections. This enables the acquisition of diverse polarizations through the rotation of the polarizer and analyzer. The apertures are used to reduce the stray light in the system. The current output of each PMT is measured by a picoamperometer and is proportional to the amout of light impinging on the photocathode.
    
In a typical scenario, the current $I_{signal}$ of the signal PMT at a specific rotated angle $P$ of the polarizer can be expressed as~\cite{ClarkJones:42,HUMLICEK20053}:
 
    \begin{equation}
    \centering
        I_{signal}\propto[1+\frac{1}{\eta}\alpha'\cos{(2A)} + \frac{1}{\eta}\beta'\sin{(2A)}]
        \label{Eq:signal_current1}
    \end{equation}
where $A$ represents the rotated angle of the analyzer, as shown in Figure~\ref{fig:ellipsometer}, $\eta=\frac{1+\xi^2}{1-\xi^2}$ where $\xi$ is the ratio of the s-mode polarized light passing through the polarizer and analyzer. To account for potential variations of $\eta$ at different rotated angles $P$ of the polarizer, we replace $\eta$ with $\eta'$ in Equation~\ref{Eq:signal_current1} by the following empirical formula:
    \begin{equation}
    \centering
        \eta'=(1+\frac{a}{b+tan^2(P-P_s)})\eta
        \label{Eq:eta_prime}
    \end{equation}
where $a$ and $b$ are two parameters that can be determined using the well-established calibration method. Thus Equation~\ref{Eq:signal_current1} can be expressed as:
    \begin{equation}
    \centering 
        I_{signal}\propto(1+\hat{\alpha}cos2A+\hat{\beta}sin2A)
        \label{Eq:signal_current2_1}
    \end{equation}
    \begin{equation}
    \centering 
        \hat{\alpha} = \frac{1}{\eta'}\alpha'
        \label{Eq:signal_current2_2}
    \end{equation}    
    \begin{equation}
    \centering 
        \hat{\beta} = \frac{1}{\eta'}\beta'
        \label{Eq:signal_current2_3}
    \end{equation}
The $\alpha'$ and $\beta'$ in Equation~\ref{Eq:signal_current1} are the two Fourier coefficients and can be determined via:
    \begin{equation}
    \centering 
        \alpha' = \frac{1}{\eta}(\alpha cos(2A_s) - \beta sin(2A_s))
        \label{Eq:alpha_prime}
    \end{equation}
    \begin{equation}
    \centering 
        \beta' = \frac{1}{\eta}(\alpha sin(2A_s) + \beta cos(2A_s))
        \label{Eq:beta_prime}
    \end{equation}
where the coefficients $\alpha$ and $\beta$ can be parameterized with the formula:

    \begin{equation}
    \centering 
        \alpha = \frac{tan^2(\psi)-tan^2(P-P_s)}{tan^2(P-P_s)+tan^2(\psi)}
        \label{Eq:alpha}
    \end{equation}

    \begin{equation}
    \centering 
        \beta = \frac{2\,tan\psi\, cos\Delta\, tan(P-P_s)}{tan^2(P-P_s)+tan^2(\psi)}
        \label{Eq:beta}
    \end{equation}
$A_s$ and $P_s$ represent the angles between the incident light plane and the p-polarization direction of the analyzer and polarizer, respectively, which need to be calibrated along with $\eta$. $\Psi$ and $\Delta$ are ellipsometric parameters~\cite{Collins20055R}, commonly utilized in commercial ellipsometers. The refractive index $n$ of the sample can be determined via:
\begin{equation}
    \centering
		n=n_{air}\sin^2(\theta_i)\sqrt{1+\tan^2\theta_i\big(\frac{1-\rho}{1+\rho}\big)}
		\label{Eq:Fresnel}
\end{equation}
where $\theta_i$ represents the angle of incidence and $\rho$ is the ratio between the two polarization components of the reflected light, given by $\rho=\frac{R_p}{R_s}$.

\begin{figure}[h]
		\centering
		\includegraphics[width=0.4\textwidth]{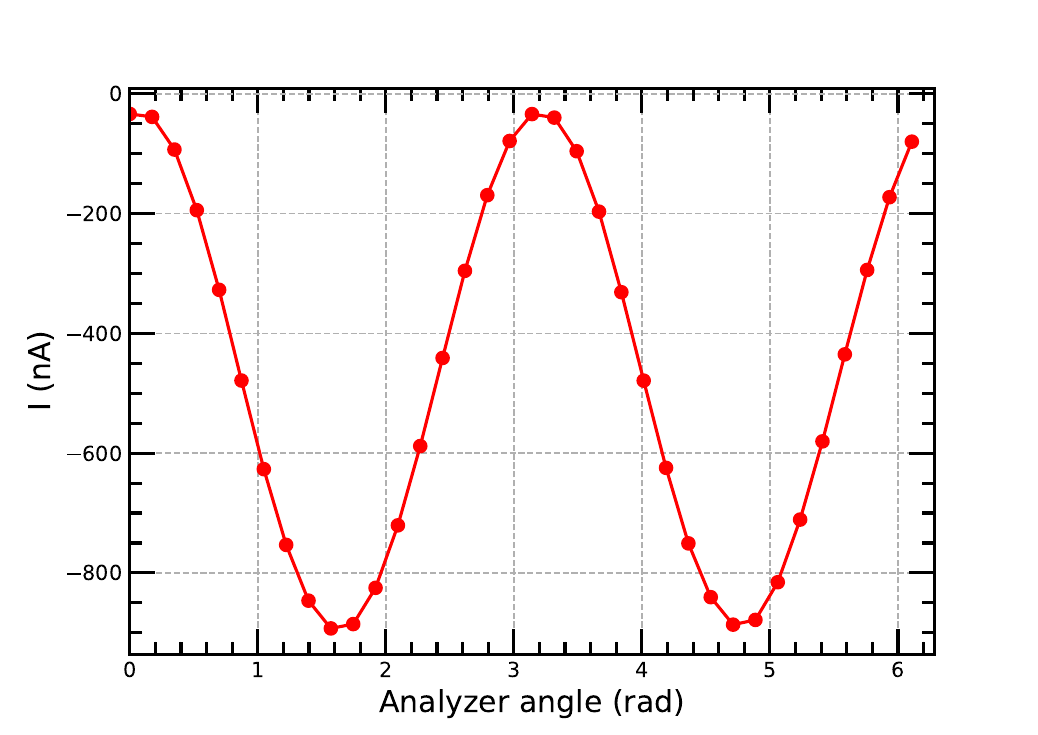}
		\caption{Signal PMT current at the polarizer angle of $P=155^\circ$ as a function of rotated angle $A$ of the analyzer.}
		\label{fig:signal_example}
	\end{figure}

\begin{figure}[h]
        \centering
        \begin{minipage}[b]{0.4\textwidth}
            \includegraphics[width=\textwidth]{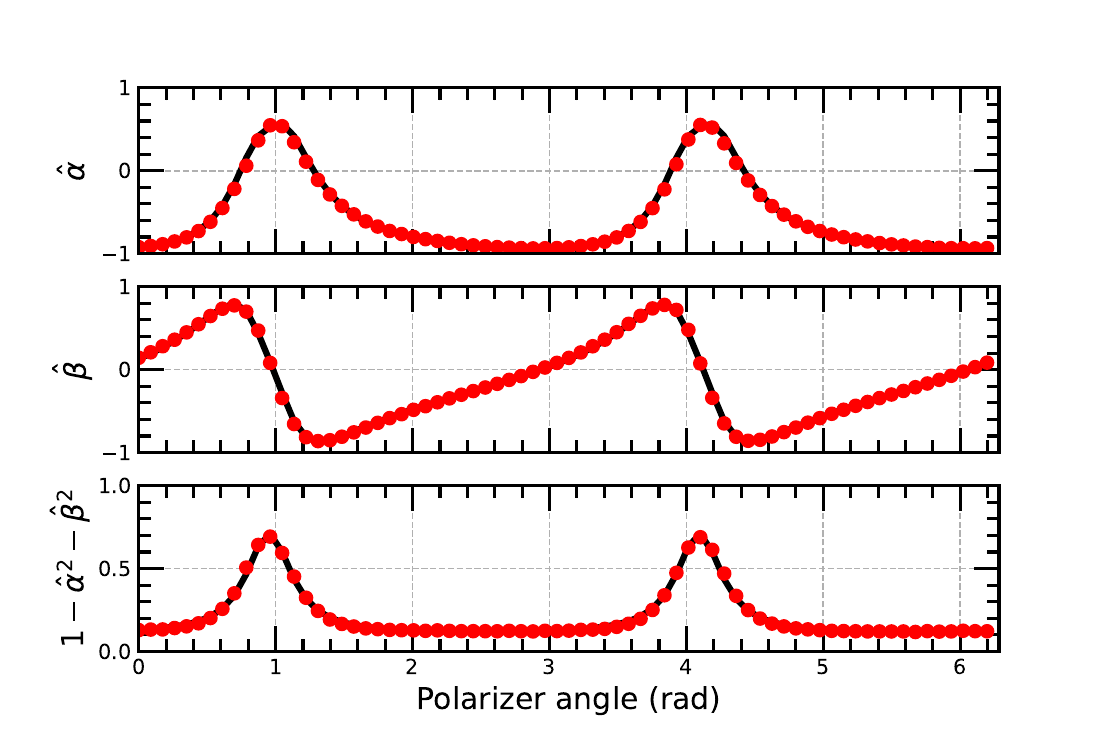}
        \end{minipage}
        \hfill
        \begin{minipage}[b]{0.4\textwidth}
            \includegraphics[width=\textwidth]{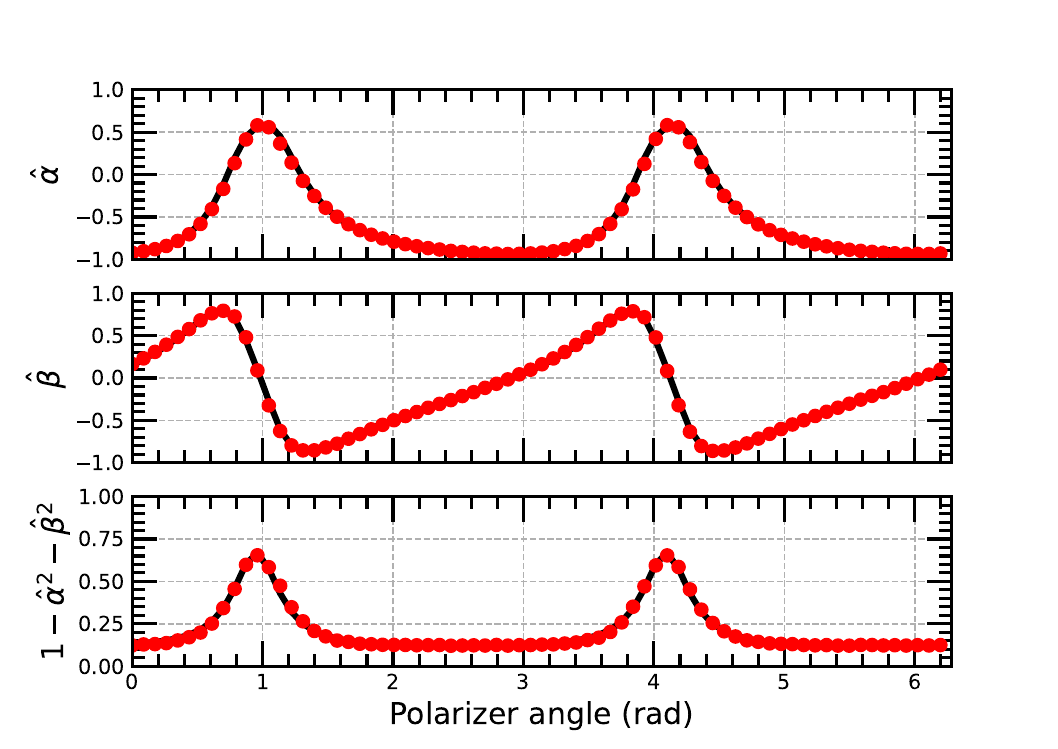}
        \end{minipage}
        \caption{Top: The Fourier coefficients of $\hat{\alpha}$, $\hat{\beta}$ and $R$ as a function of polarizer angle $P$ for PC. Bottom: The Fourier coefficients as a function of $P$ for LAB. The red dots are the obtained Fourier coefficients and the black lines represent the fitting results.}
        \label{fig:calib_fitting}
    \end{figure}

The regression calibration method is employed to obtain the aforementioned parameters of $\eta$, $A_s$ and $P_s$ in this experiment~\cite{JOHS1993395} and the calibration is performed using 400~nm light for the measured samples of LAB and PC (pseudocumene), respectively. The current of the signal PMT is measured as a function of the angle $A$ for a given angle $P$, with $P$ being scanned from 0 to $2\pi$ with an interval of $5^\circ$. The Fourier coefficients of $\hat{\alpha}$ and $\hat{\beta}$ are obtained by fitting the current curve using Equation~\ref{Eq:signal_current2_1}, and an example at $P$ = $155^\circ$ is shown in Figure~\ref{fig:signal_example}. The obtained $\hat{\alpha}$ and $\hat{\beta}$ are presented in Figure \ref{fig:calib_fitting} as a function of $P$ for pseudocumene (PC, an alternative solvent) and LAB. Subsequently $\hat{\alpha}(P)$, $\hat{\beta}(P)$ and $R(P) = 1 - \hat{\alpha}^2 - \hat{\beta}^2$ curves are simultaneously fitted to extract the calibration parameters, which are summarized in Table~\ref{tab:calib_fitting_result}. The fitting results demonstrate a good agreement with the experimental data.
 
    \begin{table}[h]
        \centering
        \begin{tabular}{c|c|c}
            \hline
            Parameters & PC & LAB \\
            \hline
            $P_s$ & 0.95096 $\pm$ 0.00010 & 0.95055 $\pm$ 0.00011 \\
            $A_s$ & 0.11809 $\pm$ 0.00007 & 0.11805 $\pm$ 0.00007 \\
            $\eta$ & 1.06492 $\pm$ 0.00006 & 1.06496 $\pm$ 0.00006 \\
            $a$ & 0.01293 $\pm$ 0.00004 & 0.01368 $\pm$ 0.00004 \\
            $b$ & 0.01727 $\pm$ 0.00008 & 0.02166 $\pm$ 0.00011 \\
            \hline
        \end{tabular}
        \caption{Summary of PC and LAB calibration parameters obtained from fitting.}
        \label{tab:calib_fitting_result}
    \end{table}

\begin{figure}[h]
		\centering
		\includegraphics[width=0.48\textwidth]{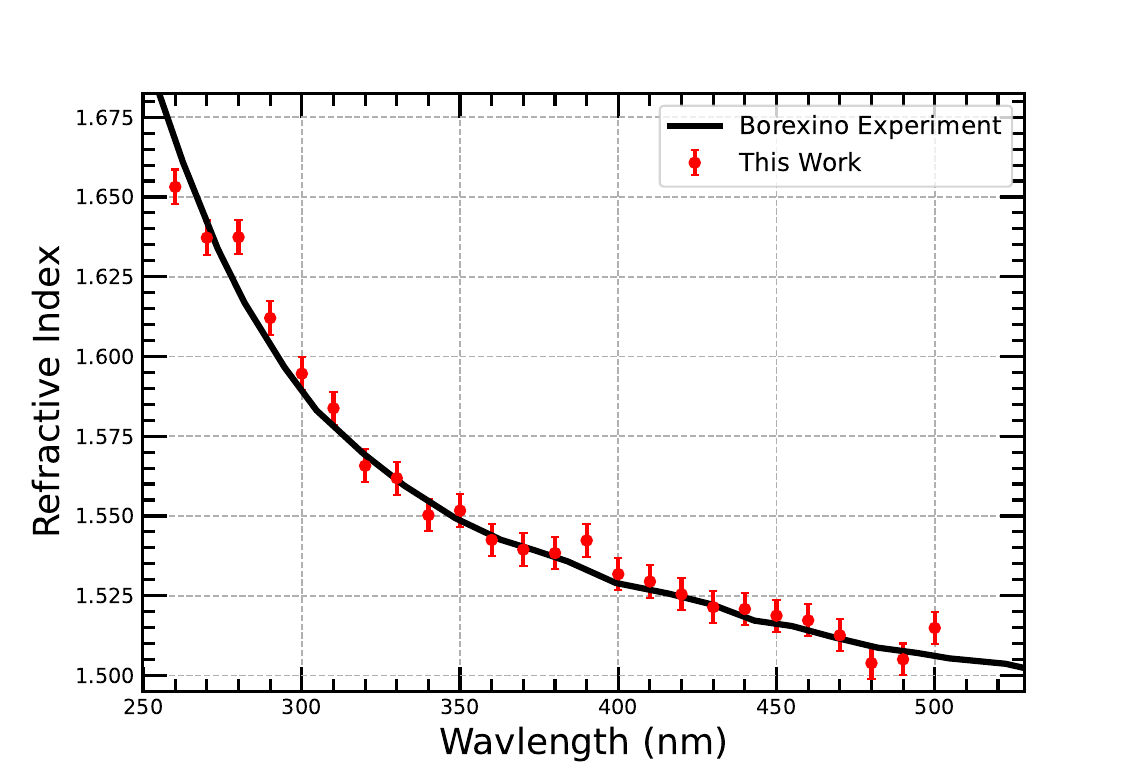}
		\caption{The refractive indices of PC as a function of wavelength. The measured values are shown as red points and the black curve is from Borexino's measurements~\cite{MarcocciPHDthesis}.}
		\label{fig:pc_rindex}
	\end{figure}

With the obtained calibration parameters, the refractive index of PC and LAB was measured within a range from 260~nm to 500~nm. Fixing the angle of the polarizer $P$ at $54^\circ$, the analyzer is rotated from 0 to $2\pi$ with an interval of $10^\circ$, and the current of the signal PMT is mapped to the analyzer rotation angle. Subsequently, the $\rho$ ratio can be determined by fitting the curve of $I_{signal}$ versus $A$ at each wavelength, with the angle of incidence measured to be $67.06^\circ$. The obtained refractive indices of LAB and PC are presented in Table~\ref{tab:rindex_result} and illustrated in Figure~\ref{fig:pc_rindex}. A good agreement with the results reported in~\cite{MarcocciPHDthesis} is evident for PC.

    \begin{table}[h]
        \centering
        \begin{tabular}{c|c|c}
        \hline
            Wavelength[nm] & Refractive index(LAB) & Refractive index(PC) \\
            \hline
            260 & 1.578 $\pm$ 0.005  & 1.653 $\pm$ 0.005  \\
            270 & 1.569 $\pm$ 0.005  & 1.637 $\pm$ 0.005  \\
            280 & 1.556 $\pm$ 0.005  & 1.637 $\pm$ 0.005  \\
            290 & 1.546 $\pm$ 0.005  & 1.612 $\pm$ 0.005  \\
            300 & 1.541 $\pm$ 0.005  & 1.595 $\pm$ 0.005  \\
            310 & 1.533 $\pm$ 0.005  & 1.584 $\pm$ 0.005  \\
            320 & 1.528 $\pm$ 0.005  & 1.566 $\pm$ 0.005  \\
            330 & 1.523 $\pm$ 0.005  & 1.562 $\pm$ 0.005  \\
            340 & 1.519 $\pm$ 0.005  & 1.550 $\pm$ 0.005  \\
            350 & 1.512 $\pm$ 0.005  & 1.552 $\pm$ 0.005  \\
            360 & 1.510 $\pm$ 0.005  & 1.542 $\pm$ 0.005  \\
            370 & 1.507 $\pm$ 0.005  & 1.539 $\pm$ 0.005  \\
            380 & 1.506 $\pm$ 0.005  & 1.538 $\pm$ 0.005  \\
            390 & 1.508 $\pm$ 0.005  & 1.542 $\pm$ 0.005  \\
            400 & 1.501 $\pm$ 0.005  & 1.532 $\pm$ 0.005  \\
            410 & 1.498 $\pm$ 0.005  & 1.529 $\pm$ 0.005  \\
            420 & 1.496 $\pm$ 0.005  & 1.525 $\pm$ 0.005  \\
            430 & 1.496 $\pm$ 0.005  & 1.521 $\pm$ 0.005  \\   
            440 & 1.494 $\pm$ 0.005  & 1.521 $\pm$ 0.005  \\
            450 & 1.497 $\pm$ 0.005  & 1.519 $\pm$ 0.005  \\
            460 & 1.492 $\pm$ 0.005  & 1.517 $\pm$ 0.005  \\
            470 & 1.489 $\pm$ 0.005  & 1.513 $\pm$ 0.005  \\    
            480 & 1.492 $\pm$ 0.005  & 1.504 $\pm$ 0.005  \\
            490 & 1.487 $\pm$ 0.005  & 1.505 $\pm$ 0.005  \\
            500 & 1.490 $\pm$ 0.005  & 1.515 $\pm$ 0.005  \\
            \hline
        \end{tabular}
        \caption{Summary of LAB and PC refractive indices obtained with ellipsometric measurements.}
        \label{tab:rindex_result}
    \end{table}
 
\subsection{Refractometric measurements}
	\label{sec:ri}
	
	To characterize the refractive index in the visible and Near Infrared spectra,  we decided to build a table-top experimental setup able to measure this important property of the scintillator. We have employed a standard refractometric technique aiming to investigate the wavelength range between 400 nm and 1100 nm. 
	
	\begin{figure}[h]
		\centering
		\includegraphics[width=0.48\textwidth]{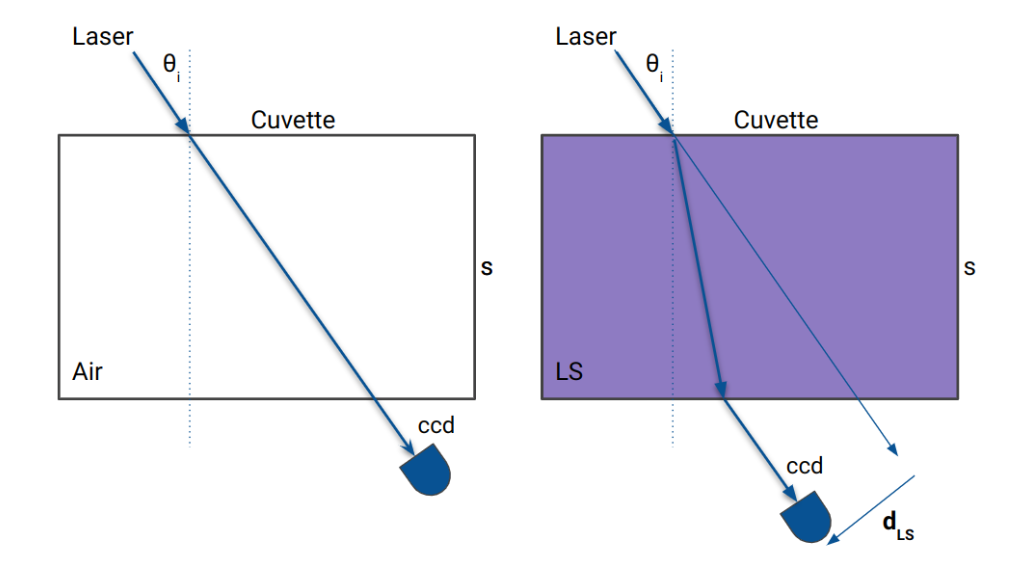}
		\caption{Schematic view of the refractometer measurement.}
		\label{fig:ri_scheme}
	\end{figure}
	
\begin{figure}[h]
		\centering
		\includegraphics[width=0.45\textwidth]{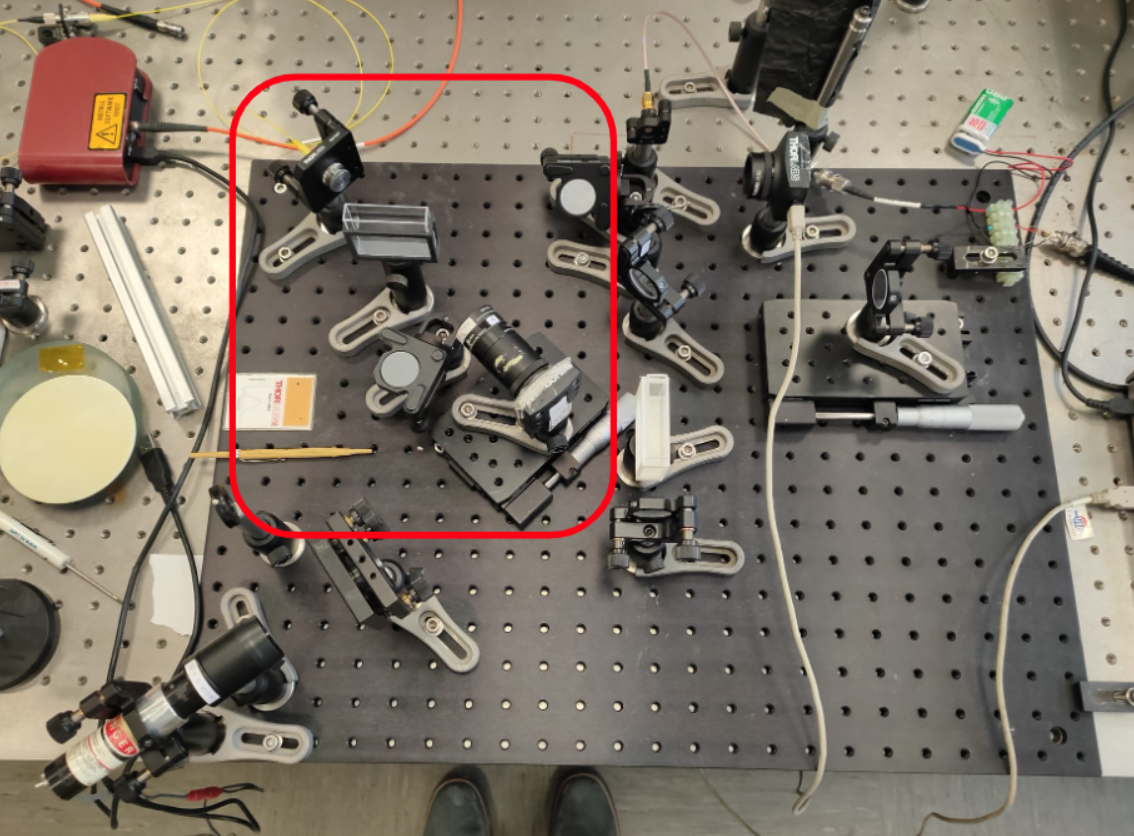}
		\caption{In the red box it is possible to see the refractometer experimental setup. In the left top edge, the single-mode optical fiber (in orange) is connected to a collimator. Then, the laser beam passes through a quartz optical cell and reaches a CDD camera.}
		\label{fig:ri_setup}
	\end{figure}

	Our refractometer consists of a collimated light source, a photosensor and a support for the sample. In particular, we used several laser sources, listed in Table~\ref{tab:table}, in the range in which the JUNO scintillator emits its fluorescence light. The LS sample is contained in $1\times5\times5$ $\mathrm{m^3}$ quartz cuvette, a DCC1545M camera (CCD) as a photosensor, a DAQ software for the acquisition, and other components for the correct propagation of light from the source to the CCD, such as collimators, optical fibers, and diffraction grating. A picture of the setup is shown in Figure~\ref{fig:ri_setup}.
	
	From the Snell law, it is possible to retrieve the refractive index by looking at the displacement of the position of a point-like source, when the beam passes through different refractive indexes:
	\begin{equation}
		\label{Eq: d}
		d = l\frac{\sin\big[\theta_i - \arcsin\big(\frac{n_a}{n_s}\sin\theta_i\big)\big]}{\cos\big[\arcsin(\frac{n_a}{n_s}\sin\theta_i)\big]}
	\end{equation}
	where $d$ is the observable displacement, $l$ is the width of the cuvette, $\theta_i$ is the incident angle of the laser beam, $n_a$ is the refractive index of the air and $n_s$ is the refractive index of the scintillator.
	\begin{table}[h]
		\caption{Laser source list}
        \vspace{0.1cm}
		\centering
		\begin{tabular}[]{ll}
  \hline
			Name     & Wavelengths [nm]\\
			\hline
			diode & 405, 670  \\
			Argon & 475, 514 \\
			Ar + Ti-sapphire     &  746, 823      \\
			Nd-YAG & 1064 \\
   \hline
		\end{tabular}
		\label{tab:table}
	\end{table}

	For each measurement, we checked that the laser beam arrived perpendicular to the CCD camera, within less than one degree precision which implies an error order of $10^{-4}$ on the displacement measurement, negligible for our purpose.	Hence, two possible factors limiting the precision of the measurements are the uncertainty on the incident angle on the cuvette ($\theta_i$) and the uncertainty on the displacement. The former can be calibrated using a material, such as water, with a well-known refractive index. Therefore, the main source of error is the uncertainty of the displacement.
	
	
	In order to avoid contamination between water and liquid scintillator in the optical path of the laser beam in the cuvette, we measured the displacement filling the cuvette first with the scintillator and then with distilled water. Since the density of the LAB is lower than the one of the water,
	even if small LAB residuals remain in the cell, it does not spoil the measurements since they will float on top of water. If water was used first instead, some drops of water could stay wrapped in the liquid scintillator causing some unwanted reflections. We have tested and validated our procedure by measuring some "well-known" refractive indexes, like Ethanol, finding it in agreement with the value in literature within a 0.1$\%$ accuracy.  The results obtained on the LAB with this technique are shown in Table~\ref{tab:Milano_results} and depicted in Figure~\ref{fig:comb} with red dots.
	
	
	\begin{table}[h]
		\centering
		\caption{Experimental results of LAB refractive indexes by the refractometer measurements}
  \vspace{0.1cm}
		\begin{tabular}{c|c|c}
         \hline
            Wavelength [nm] & Refractive index & Error\\
			\hline
			405 & 1.505 & 0.007\\
            476 & 1.486 & 0.007\\
            514 & 1.490 & 0.007\\
            633 & 1.473 & 0.003\\
            670 & 1.478 & 0.007\\
            746 & 1.473 & 0.007\\
            823 & 1.469 & 0.007\\
            1064 & 1.468 & 0.007\\
            \hline
		\end{tabular}
		\label{tab:Milano_results}
	\end{table}	

\subsection{Results and analysis}
	Using these two techniques, it was possible to measure the refractive index in the range between $260\,\mathrm{nm}$ and $1064\,\mathrm{nm}$ reaching a precision ranging from $\sim1\%$ to $\sim0.2\%$. Using the Sellmeier law, we can verify the trend of the data. 
	\begin{equation}
	n^2(\lambda) = 1 - \frac{B}{1-C/\lambda^2}
	\label{Eq:Sell}
	\end{equation}
	
\begin{figure}[h]
		\centering
		\includegraphics[width=0.5\textwidth]{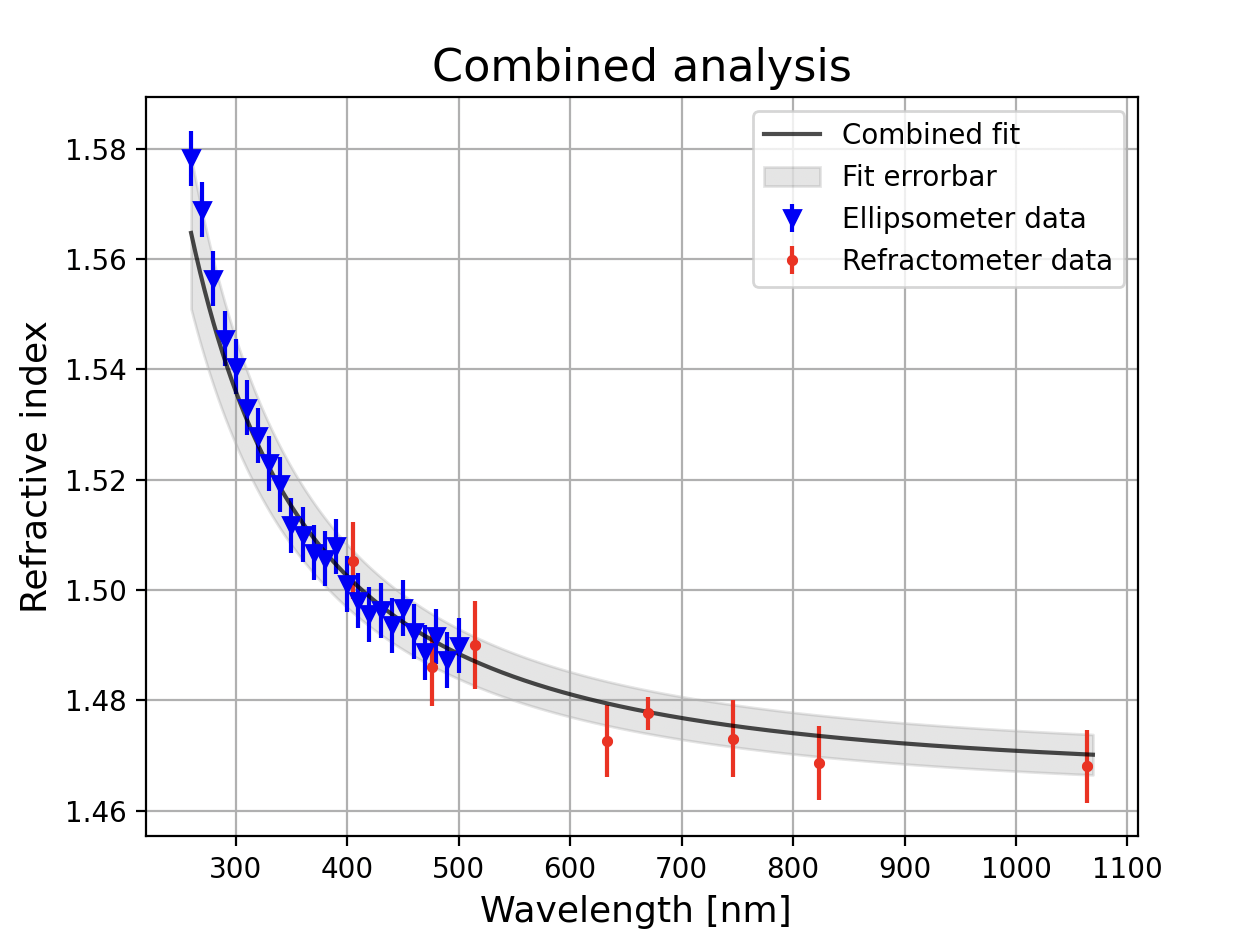}
		\caption{Combined analysis of the two datasets. As it is possible to see both the fits are in the error band of the global fit.}
		\label{fig:comb}
	\end{figure}

	The global fit of the two datasets is shown in Figure~\ref{fig:comb}. As it is possible to see, the data of the two datasets (ellipsometer in blue, refractometer in red) are compatible in the region from $400\,\mathrm{nm}$ and $500\,\mathrm{nm}$ and all the experimental points are compatible with the fit. The results on the Sellmeier fit are shown in Table~\ref{tab:tot}. Combining the two methods, we increase the accuracy and precision of this crucial parameter of the liquid scintillator.
	\begin{table}[h]
		\centering
		\caption{Summary of the fit results applying the Sellmeier law on the two datasets.}
        \vspace{0.1cm}
		\begin{tabular}{c|c|c|c|c}
  \hline
			Dataset & B & $\sigma_B$ & C [nm$^2$]& $\sigma_C$ [nm$^2$]\\
			\hline
			Combined & 1.147 & 0.003 & 14.1$ \times 10^3$ & 0.5 $ \times 10^3$\\ 
   \hline
		\end{tabular}
		\label{tab:tot}
	\end{table}

\section{Group velocity of the JUNO liquid scintillator}
The refractive index impacts event reconstruction in JUNO mostly for what concerns position reconstruction. We measured directly the group velocity in the liquid scintillator to check our measurement of the refractive index.  Indeed, for a dispersive medium, it is possible to write the phase velocity as the ratio between the speed of light and the refractive index. Hence the group velocity is described by: 
	\begin{equation}
		v_g(\lambda) = \frac{c}{n(\lambda)}\bigg(1-\frac{\lambda}{n(\lambda)}\frac{dn(\lambda)}{d\lambda}\bigg)^{-1}
		\label{vg}
	\end{equation}
	To perform this measurement, we used a standard Michelson interferometer, which is a device allowing us to study the interference of two beams. In particular, a laser beam passes through a semi-reflective mirror (a beam splitter), which splits it along two arms. One of these arms ends with a plane mirror reflecting back the light. The other arm has a 5 cm cuvette in which the beam passes reaching a mirror in order to be reflected back. Then the two beams are recombined and seen by a fast photosensor which acquires the data.
	\begin{figure}[h]
		\centering
		\includegraphics[width=0.35\textwidth]{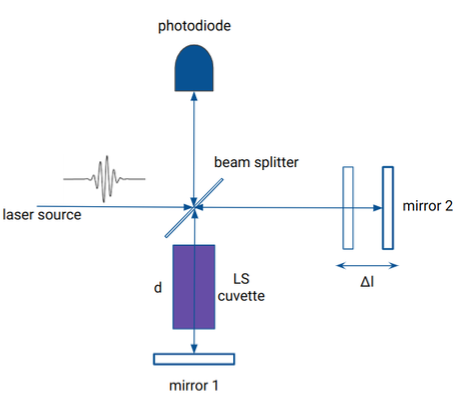}
		\caption{Schematic view of group velocity measurement performed using a Michelson interferometer.}
		\label{fig:int}
	\end{figure}
	
	 Inserting a sample in the cell along the optical path, it is possible to see a shift in the position of the interference fringes. Measuring this shift via a micrometric screw, which moves the mirror, it is possible to retrieve the group velocity. A scheme of the setup is shown in Figure~\ref{fig:int}.
	 
	 	\begin{figure}[h]
	 	\centering
	 	\includegraphics[width=0.45\textwidth]{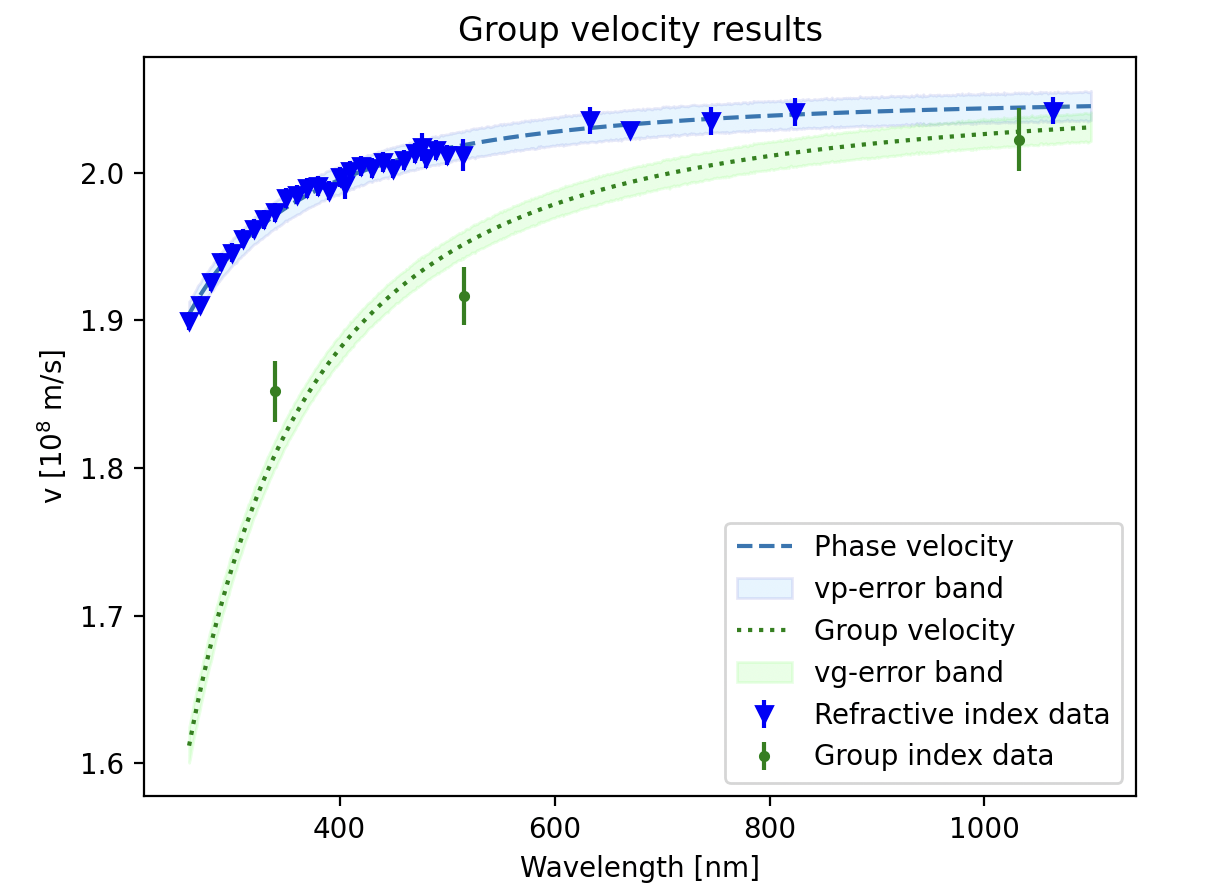}
	 	\caption{In blue it is possible to see the phase velocity curve determined by the refractive index measurements. The group velocity measurements are depicted in green. The green line instead represents the group velocity curve calculated from the refractive index measurements using the formula \ref{vg}.}
	 	\label{fig:vg}
	 \end{figure}

	 We measured the group velocity using two laser sources reaching three different wavelengths: 340, 516, and 1036 nm. The results are shown in Figure~\ref{fig:vg} overlaying on the predicted group velocity curve obtained by using the formula in Equation~\ref{vg} and the Sellmeier fit on the refractive index measurements. The data, shown in Table~\ref{tab:vg}, follow the predicted law in the full spectrum at $2\,\sigma$ level.
	 
	\begin{table}[h!]
		\centering
		\caption{Results of the group velocity measurements in a fraction of the light speed.}
			\begin{tabular}{c|c}
   \hline
		Wavelength [nm]& $v_g/c$\\
  \hline
        340 &  0.6178$\pm$ 0.0069\\
        516 & 0.6392$\pm$0.0066\\
        1036 & 0.6746$\pm$0.0071 \\
        \hline
	\end{tabular}
	\label{tab:vg}
	\end{table}

\section{Conclusion}
	JUNO will be a very large detector based on 20,000 tons of organic liquid scintillator. To reach its physics goals, the experiment needs an accurate Monte Carlo code with the best possible description of the liquid scintillator. Different groups of the collaboration started programs to characterize the liquid scintillator accurately. In this work, we summarize the experimental results on the refractive index obtained by two different groups of collaboration in order to extend the results in the largest optical range possible. These results on the refractive index will help to better understand the Cherenkov contribution in the liquid scintillator and to improve the position reconstruction in the JUNO detector. Furthermore, the group velocity measurements support the measurements of the refractive index and allows the best measurement of the light speed in the JUNO liquid scintillator.
	
\section*{Acknowledgments}
The IHEP group gratefully acknowledges support from the National Natural Science Foundation of China (NSFC) under grant No. 12375196. The Milan group would like to thank Professor Cialdi of the University of Milan for the time and the knowledge shared with us for both the refractometer and the interferometer. Many thanks also to Milan University for the financial support needed to build the experimental setup. 

\bibliographystyle{elsarticle-num}

\bibliography{example.bib}









\end{document}